\documentstyle[twocolumn,aps,prl,floats,epsfig]{revtex}

\def\YBCO{YBa$_2$Cu$_4$O$_8$}
\def\YBC3O#1{YBa$_2$Cu$_3$O$_{#1}$}
\def\Gd{Gd$^{3+}$}
\def\CuO{CuO$_2$}
\def\Hc#1{H_{c#1}}
\def\GdK{{}^{\rm Gd}K}
\def\GdA{{}^{\rm Gd}A}
\def\chis{\chi_{\rm s}}

\def\comment#1{}

\begin{document}

\draft

\wideabs{
\title{Magnetic Field Induced Low-Energy Spin Excitations in
\YBCO\ Measured by High Field \Gd\ ESR}
\author{Titusz Feh\'er$^1$, Andr\'as J\'anossy$^{1,2}$,
G\'abor Oszl\'anyi$^{1,2}$, Ferenc Simon$^1$,\\
Bogdan Dabrowski$^3$, Piotr W.~Klamut$^3$, Mladen Horvati\'c$^4$, and
Grant V.~M.~Williams$^5$}

\address{$^1$Institute of Physics, Technical University of Budapest,
P.~O.\ Box 91, H-1521 Budapest, Hungary}
\address{$^2$Research Institute for Solid State Physics, P.~O.\ Box 49,
H-1525 Budapest Hungary}
\address{$^3$Department of Physics, Northern Illinois University, De
Kalb, Illinois 60115, USA}
\address{$^4$Grenoble High Magnetic Field Laboratory, CNRS and MPI-FKF,
BP 166, F-38042 Grenoble Cedex 9, France}
\address{$^5$New Zealand Institute for Industrial Research, P.~O.\ Box
31310, Lower Hutt, New Zealand}

\date{\today}

\maketitle

\begin{abstract}
We have measured the spin susceptibility of the underdoped high
temperature superconductor, \YBCO\ by \Gd\ electron spin resonance in
single crystals and aligned powders at several magnetic fields between
$3$ and $15.4\rm\,T$. At low temperatures and high fields, the spin
susceptibility of the \CuO\ planes is enhanced slightly in the
$B\parallel c$ orientation with respect to the $B\perp c$ orientation.
The enhancement in an applied field of $15.4\rm\,T$
($\approx0.15\,\Hc2$) at $16\rm\,K$ ($0.2\,T_c$) is approximately 10
percent of the susceptibility measured at $T_c$. Such a small magnitude
suggests that the second critical field of superconductivity,
$\Hc2\approx100\rm\,T$, would {\em not} suppress the pseudogap. This
work demonstrates the potential of high field ESR in single crystals for
studying high $T_c$ superconductors.
\end{abstract}


} 

The structure of vortex lines in high temperature superconductors (HTSCs)
may shed light on the microscopic mechanism of superconductivity. The
earliest $d$-wave pairing theories implied that the zero temperature spin
susceptibility should scale with magnetic field as $(B/\Hc2)^{1/2}$\cite
{osVolovik,osMaki}\ in contrast to the linear dependence in an $s$-wave
superconductor. A rigorous solution of the Bogoliubov--de Gennes equations
in the mixed state\cite{QPtheory} reveals that quasiparticles in $d$-wave
superconductors are not bound to vortices, and predict a space
averaged density of states (DOS) $N(0)\propto(B/\Hc2)^\gamma$, where
$0.4\le\gamma\le1$ depending on the temperature and the model used
\cite{KopninVolovik,Ichioka_0.41}. A number of experiments were targeted
at the field dependence of the DOS in \YBC3O{7-\delta}.  Scanning
tunneling microscopy\cite{MaggioaprileSTM} resolves the DOS at and around
vortices, but is unsuitable to measure the total DOS\cite{c-tunneling}. Bulk
experiments on nearly optimally doped \YBC3O{7-\delta}\ include heat
capacity\cite{MolerPRB}, infrared transmission\cite{Karrai}, thermal
conductivity\cite{QPtransport}, and high field muon spin
rotation\cite{Sonier_highfieldmuSR}.
Most of these experiments, together with early NMR on planar copper and
oxygen\cite{osNMR}, indicated some increase in the DOS with magnetic
field but were inconclusive about its magnitude.

Underdoped HTSCs---which also possess a pseudogap in the normal state---are
of special interest, as it is unclear how the pseudogap is related to the
$d$-wave superconducting gap. An indication whether high magnetic fields
suppress the pseudogap or not may contribute to understanding its nature.
Also, most theoretical calculations of the quasiparticle spectra in HTSCs
depend on the assumption that Landau's theory of the Fermi-liquid is
applicable below $T_c$, which is still to be justified by experiments.
A recent $^{63}$Cu NMR study on \YBCO\ at high fields\cite{ZhengClarkNMR}
reports a sizeable field induced DOS for $B$ in the $(a,b)$ plane, implying
a large enhancement for $B\parallel c$. In contrast, we show that for
$B\parallel c$ the DOS enhancement is small and incompatible with a
suppression of the pseudogap at $B=\Hc2$.

In this Letter, we report the spin susceptibility of \YBCO\ obtained
by high field electron spin resonance (ESR) spectroscopy of \Gd\ in
Gd:\YBCO.  \Gd\ is a non-perturbing probe of the \CuO\ spin
polarization\cite{gdCooper}. We search for low-energy spin excitations
induced by high magnetic fields in the superconducting phase.
Our goal is to determine the field dependence of the anisotropy of the
susceptibility, $\chis^c-\chis^{ab}$, from the \Gd\ ESR shift. The
anisotropy is a measure of the magnetic field induced DOS along $c$ since
$\Hc2$ in the $(a,b)$ plane is several times larger than in the $c$
direction.  We measure both oriented powder (OP) and untwinned single
crystal (SX) samples at several temperatures, frequencies, and magnetic
field orientations. SX data are used to determine the zero field splitting
(ZFS) parameters in the spin Hamiltonian of \Gd\ in Gd:\YBCO\ with
precision. This allows us to model powder spectra and measure shifts in the
OP sample in both $B\parallel c$ and $B\perp c$ orientations with high
accuracy. Diamagnetic effects inhibit SX measurements at low temperatures.


The experiments were carried out on the high field ESR spectrometers in
Budapest and Grenoble. In Budapest, a quartz oscillator stabilized
Gunn-diode was used as a mm-wave source at $f$=75, 150, and $225\rm\,GHz$.
In Grenoble, we used the same setup with a Gunn-oscillator at 95, 190, and
285$\rm\,GHz$ and an optically pumped far infrared laser at 349 and
$429\rm\,GHz$, and a sweepable 17 Tesla NMR magnet. The frequency of
$429\rm\,GHz$ corresponds to $15.4\rm\,T$ central \Gd\ resonance field.  The
radiation is transmitted through the sample in an oversized waveguide and no
cavity is used. We detect the derivative of the absorption with respect to
magnetic field.  The magnetic field is calibrated by a reference sample,
BDPA \comment{($g_{\rm BDPA}=2.00359$)}\cite{gdCooper}, at each sweep. The
powder sample Gd:\YBCO\ was sintered by a standard solid state reaction,
with 1\% of Y substituted by Gd. The powder with a characteristic grain size
of $5\rm\,\mu m$ was mixed with epoxy resin, and aligned in a magnetic
field. SX samples of typical dimensions $1\times 1\times 0.2\rm\,mm^3$ were
grown by self-flux method at $1100^\circ\rm C$ in $600\rm\,atm$~of O$_2$.
$T_c=80\rm\,K$ in both the SX and the OP samples.

The exchange induced shift of the \Gd\ ESR lines is similar to the $^{89}$Y
NMR Knight shift and yields the spin susceptibility of the \CuO\ planes. We
define the \Gd\ ESR shift, in analogy with the NMR notation as

\begin{equation}
\GdK^\alpha(B_0,\,T)=-(B_m^\alpha(B_0,\,T)-B_0)/B_0
\label{eq:defK}
\end{equation}

\noindent with $B_0=hf/g_0\mu_B$, where $f$ is the microwave frequency
and the arbitrary zero of $\GdK$ is defined by $g_0=1.9901$.
$B_m^\alpha$ is the measured resonance field as deduced from the
measured positions of the \Gd\ lines (after correcting for the ZFS,
i.e., $B_m^\alpha$ would be the resonance field of a similar probe with
no ZFS). Then

\begin{equation}
\GdK^\alpha(B_0,\,T) =
   K_0^\alpha+\GdA\chis^\alpha(B_0,\,T)-B_{\rm dia}^\alpha(B_0,\,T)/B_0,
\label{eq:linepos}
\end{equation}

\noindent where $K_0^\alpha=(g_{\rm Gd}^\alpha-g_0)/g_0$ plays
the role of the NMR chemical shift. We denote by $g_{\rm Gd}^\alpha$
the ``pure'' $g$-factor of \Gd\ in Gd:\YBCO\ from which the exchange
with the \CuO\ spins is eliminated. $g_{\rm Gd}^\alpha$ (thus $K_0^\alpha$) is
independent of magnetic field and temperature\cite{afdomains}, and its
anisotropy is small. The spin susceptibility of the \CuO\ planes is
defined by $\chis^\alpha(B,\,T)=M_{\rm s}^\alpha(B,\,T)/B$ where $M_{\rm
s}^\alpha$ is the \CuO\ spin magnetization. The \Gd\ shift due to the
electronic exchange interaction between \Gd\ and \CuO\ is linked to the
susceptibility through the constant
$\GdA\approx-15\rm\,mole/emu$\cite{gdlinprl} (in analogy with the NMR
hyperfine constant).
We neglect the anisotropy of $\GdA$, since a comparison of the anisotropy of
the shifts in the normal state \YBC3O7\cite{gdCooper} and antiferromagnetic
\YBC3O6\cite{afdomains} shows that the anisotropy of the product
$\GdA\chis^\alpha$ is approximately 5\%. $B_{\rm dia}^\alpha$ is the bulk
demagnetizing
field of the
supercurrents in the crystallites. In this work we shall estimate the vortex
contribution to $\chis^c$ at high magnetic fields from
$\Delta B/B_0
=(\GdK^c-\GdK^{ab})-(K_0^c-K_0^{ab})
=\GdA(\chis^c-\chis^{ab})-(B_{\rm dia}^c-B_{\rm dia}^{ab})/B_0$.
Here $\GdK^\alpha$ is measured directly (Eq.~(\ref{eq:defK})), and
$K_0^c-K_0^{ab}$ is determined from the high field data at high
temperatures. Unlike the $^{89}$Y case, $B_{\rm dia}^\alpha$ is not
prohibitively large since $\GdA$ is ten times greater
than the $^{89}$Y hyperfine constant, $^{89}A$.

\begin{figure}[t]
\noindent
\includegraphics[width=\columnwidth]{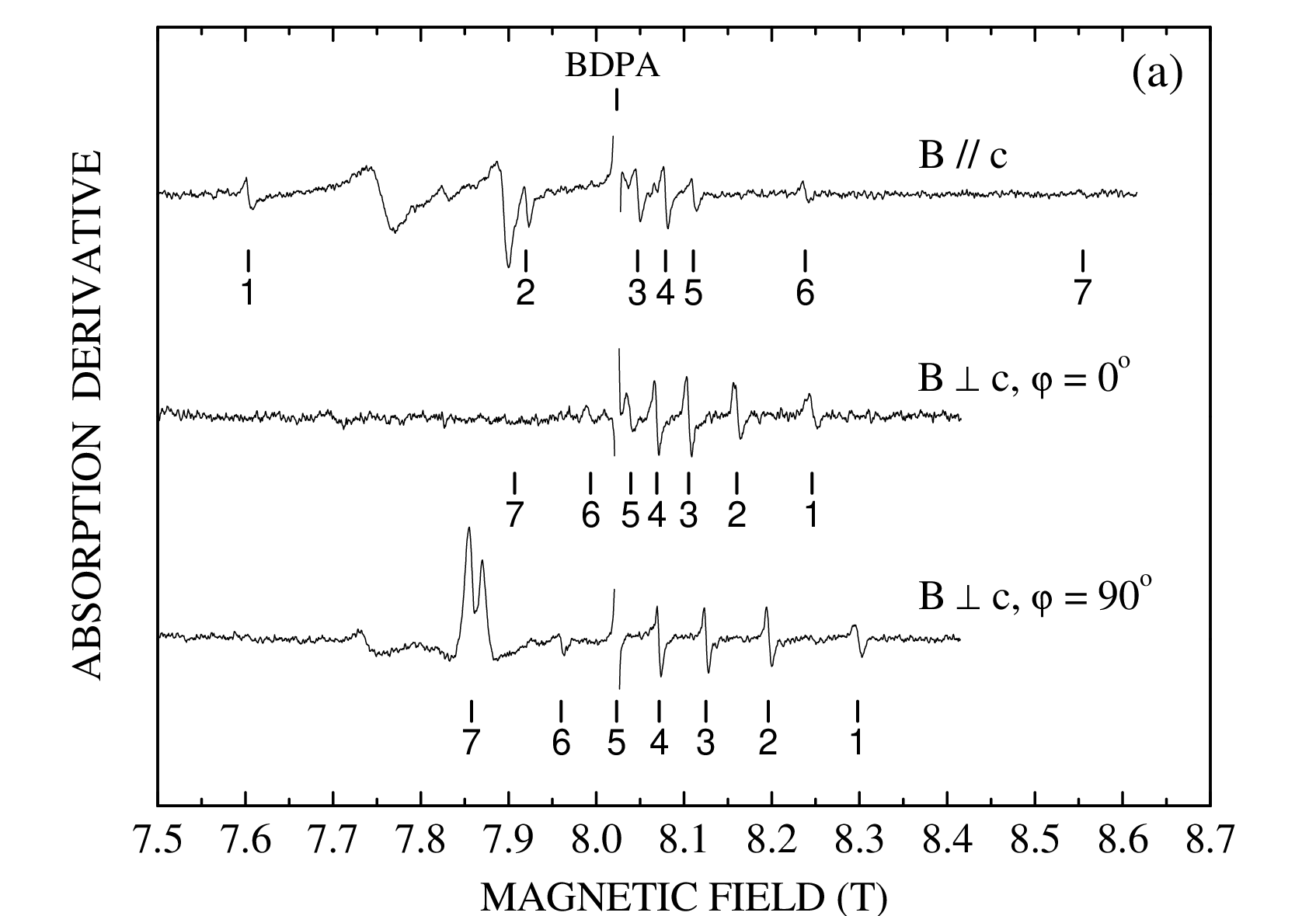}
\includegraphics[width=\columnwidth]{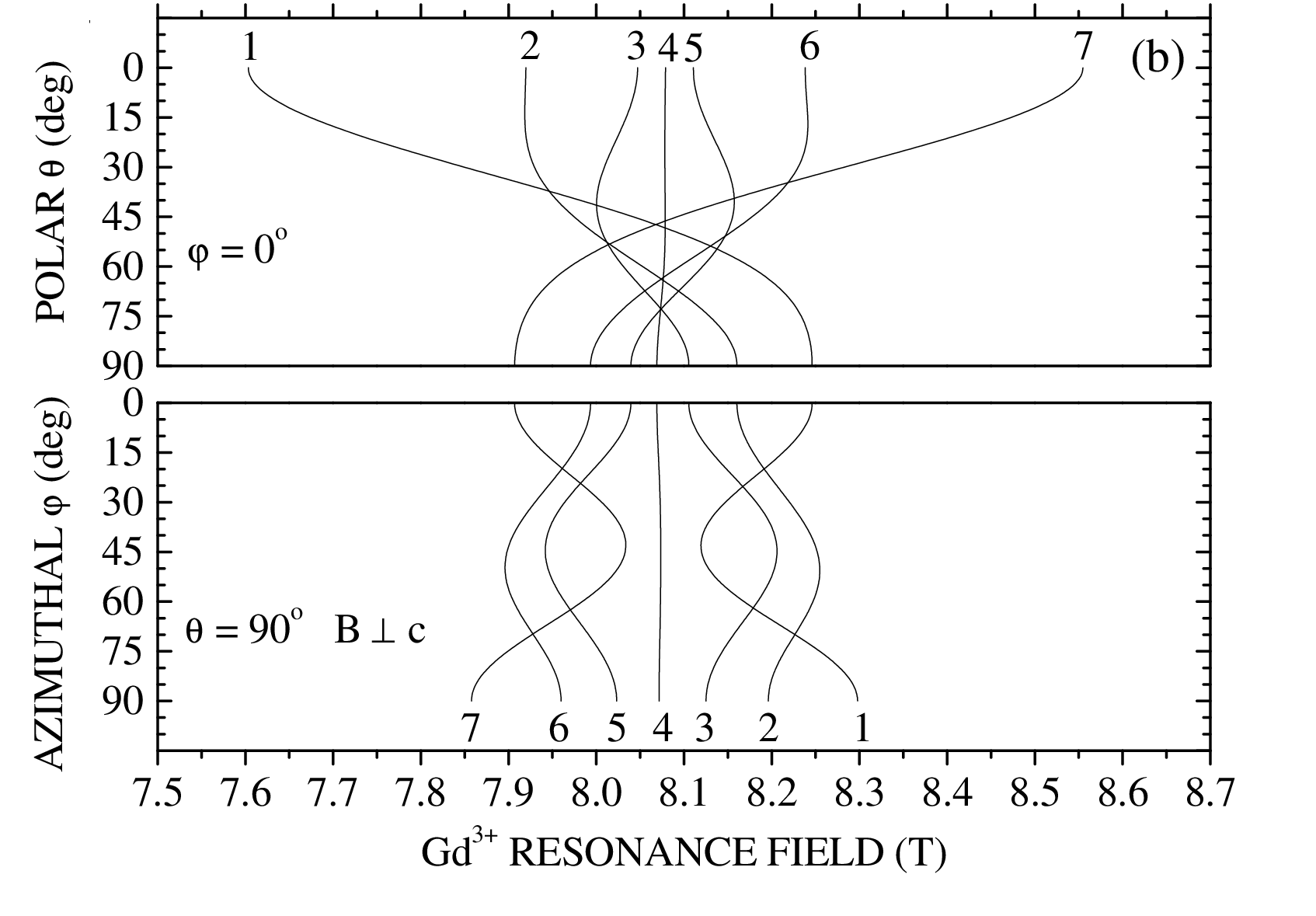}
\caption{\label{f:Dab}
(a) \Gd\ ESR spectra of a single crystal Gd:\YBCO\ taken with magnetic
field along the three crystalline axes at $T=40\rm\,K$ and
$f=225\rm\,GHz$.  Neither twinning nor mosaicity is observed. The field
reference BDPA is cut out for clarity. Numbers $1,\,\ldots,\,7$ denote
transitions $|-7/2\rangle\rightarrow|-5/2\rangle,\, \ldots,\,
|+5/2\rangle\rightarrow|+7/2\rangle$, respectively. Bars ($|$) show
simulated line positions. We do not know whether $\varphi=0^\circ$ belongs
to $B\parallel a$ or~$b$.  (b) Positions of the lines calculated from the
spin Hamiltonian of \Gd\  as a function of magnetic field orientation at
$225\rm\,GHz$.
}
\end{figure}

The temperature independent ZFS parameters of the nearly pure S state
$S=7/2$ \Gd\ ion were obtained from the SX data. The ZFS of \Gd\ is smaller
than the Zeeman splitting in the fields employed and the spectrum consists
of the 7 allowed fine structure transitions, whose relative positions depend
little on the ESR frequency but are sensitive to the orientation of the
sample with respect to magnetic field. Spectra of single crystals were
recorded in the temperature range 30--$70\rm\,K$. Fig.~\ref{f:Dab}(a) shows
the 225 GHz spectra of an untwinned SX with magnetic field along the three
crystalline axes at $40\rm\,K$. The principal axes of the spin
Hamiltonian of the \Gd\ in its orthorhombic environment coincide with the
crystalline axes. At $40\rm\,K$ \Gd\ ESR lines are narrow and about the
same width for all transitions showing that there are no strains or
inhomogeneities in the high quality crystals. Although the small
microwave penetration depth necessitates a few hours of averaging for
each spectrum, the potential to use \Gd\ ESR in small single crystals is
clear, e.g., to measure internal fields around impurity atoms. The use
of resonant cavities renders conventional X-band ESR spectrometers
unsuitable to study the superconducting state. Although there are
several reports on \Gd\ ESR in perovskites, we do not know of any other
report on the ESR of HTSC single crystals below $T_c$. Measuring the
positions of the lines with magnetic fields in several orientations and
at three frequencies allows us to fit the ZFS parameters of
\Gd\cite{cfparams}. Details of the spin Hamiltonian in a similar
compound, \YBC3O{6+x}, were published elsewhere\cite{afdomains}. The
spin Hamiltonian describes the SX spectra well, the difference between
the calculated and measured line positions is always less than the line
width.  The simulated line positions as a function of orientation used
for evaluating the OP spectra are indicated in Fig.~\ref{f:Dab}(b).

\begin{figure}[t]
\noindent
\includegraphics[width=\columnwidth]{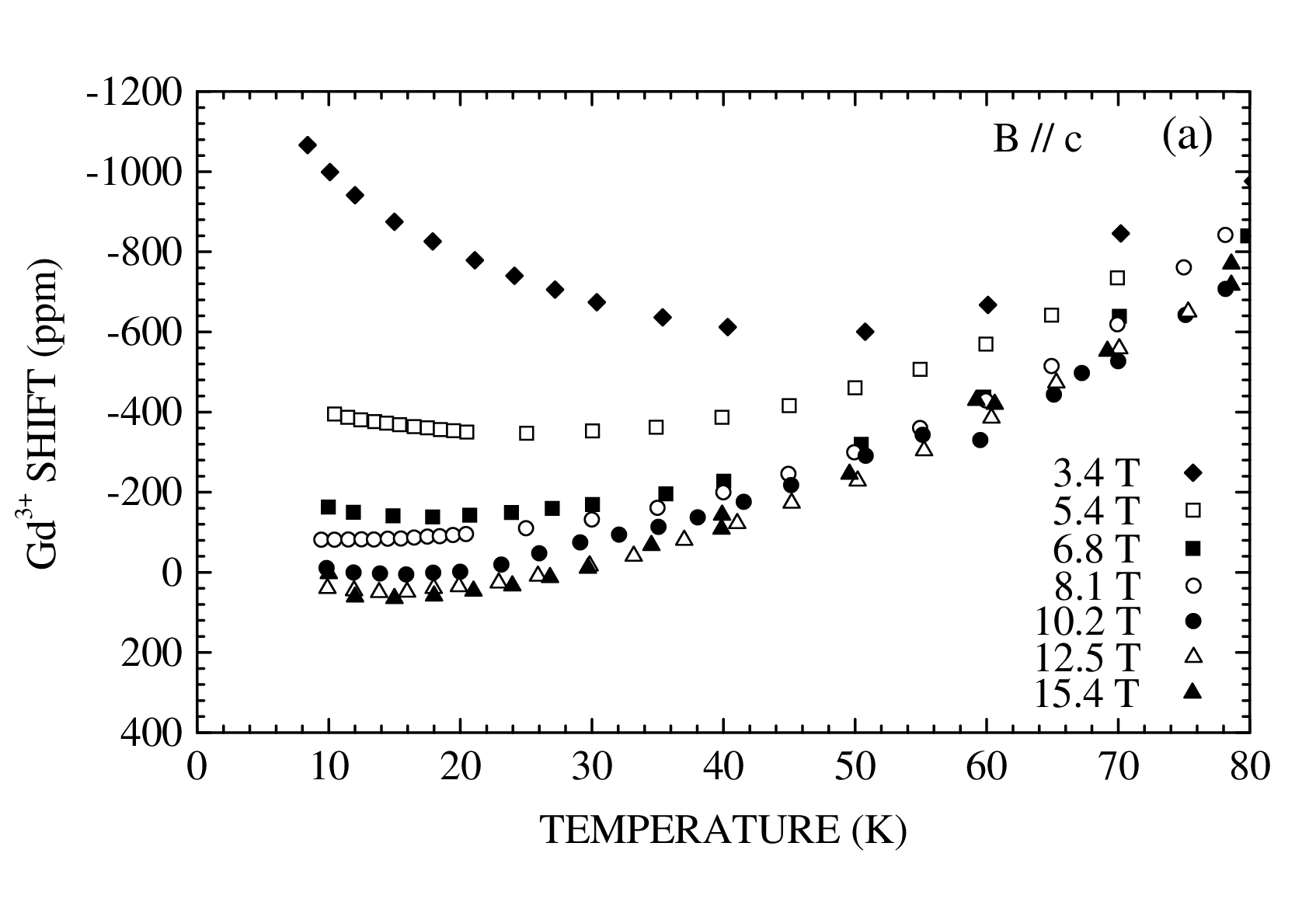}
\includegraphics[width=\columnwidth]{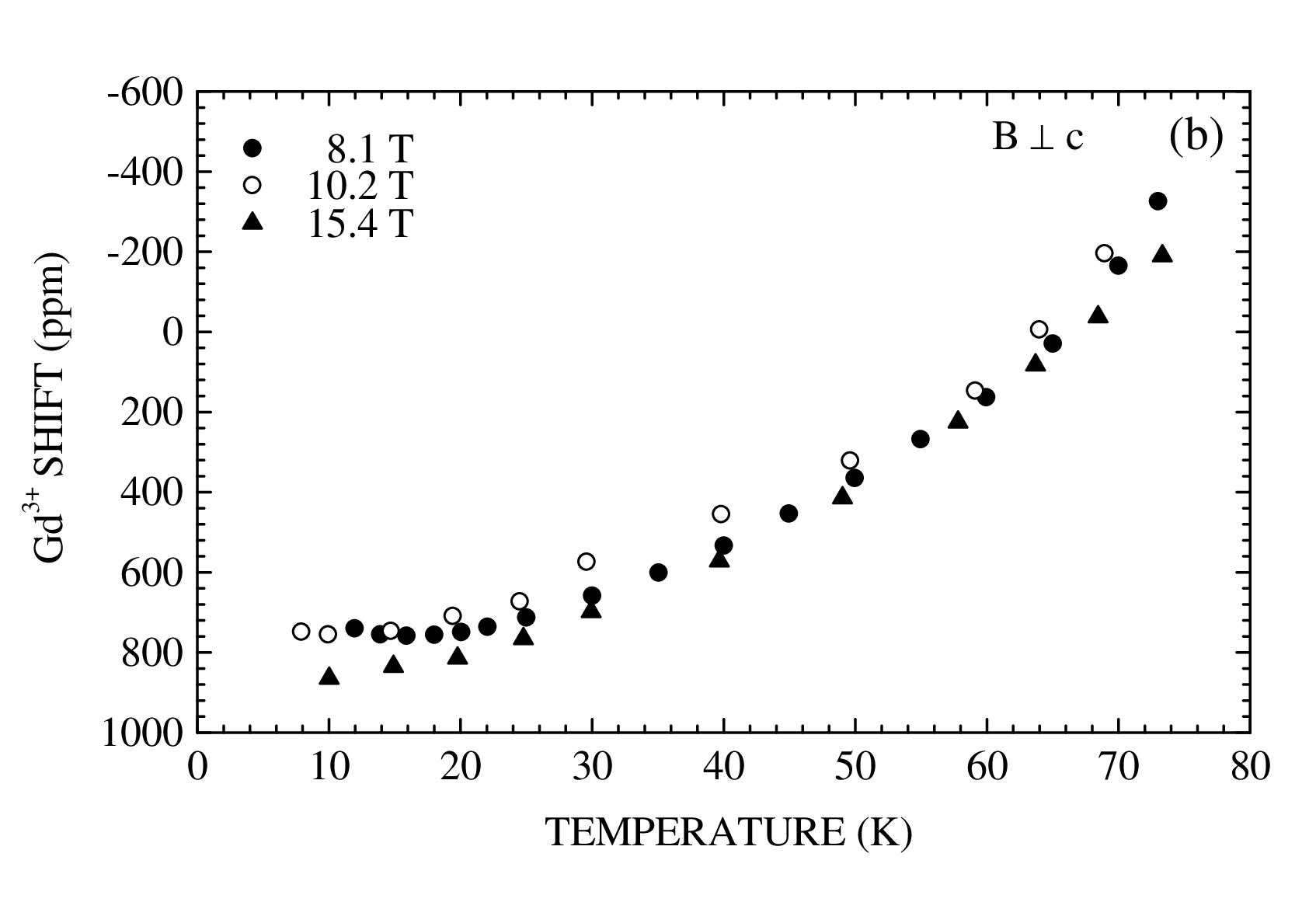}
\caption{\label{f:shift}
Normalized \Gd\ ESR shifts, $\GdK^\alpha(B,\,T)$, in the OP sample
as a function of temperature at several magnetic fields. (a) $\alpha=c$,
(b) $\alpha=ab$.
}
\end{figure}

Once the ZFS parameters are obtained from the SX, the \Gd\ shifts,
$\GdK^\alpha(B,\,T)$, may be deduced with high accuracy from the
OP spectra. To obtain the increase of $\chis^c$ with magnetic field we use
$\GdK^\alpha$ measured at several $T$ and $B$ in both $\alpha=c$ and $ab$
directions as shown in Fig.~\ref{f:shift}.  These data are uncorrected for
reversible diamagnetic fields, while a small irreversibility in $B_{\rm
dia}^\alpha$ at low $T$ and low $B$ is eliminated by averaging $\GdK^\alpha$
of spectra taken with field swept up and down at the same
temperature\cite{gdlinprl}.
At low $T$ and high $B$ the central lines fade away, we could therefore
measure the shift in $15.4\rm\,T$ reliably only above $15\rm\,K$. The
dependence of the normalized shift on the resonance frequency
(Fig.~\ref{f:shift}) is in a large part caused by diamagnetic shielding.
However, as shown below, in the $c$ direction at high fields there is also
some contribution from the vortex spin susceptibility.

We assume that only $\chis^c$ has a measurable field dependence
at the fields employed, and that of $\chis^{ab}$ may be
neglected. This is consistent with our main result below that the field
dependence of $\chis^c$ is weak, since even if there is some field
dependence of $\chis^{ab}$, it is significantly smaller than that of
$\chis^c$, because of the much higher $\Hc2^{(ab)}$. 
Diamagnetic corrections to $B_m^{ab}$ are also smaller than to $B_m^c$ as
$\lambda_c\gg\lambda_{ab}$ holds for the penetration depths. The measured
field dependence of $\GdK^{ab}$ is weak (Fig.~\ref{f:shift}(b)), and at
$15.4\rm\,T$ we neglect the diamagnetic correction. At $15.4\rm\,T$ the raw
$ab$ data are already a good approximation of the spin susceptibility.
Indeed, the shift at $15.4\rm\,T$ decreases with temperature at low $T$ and
the slope is close to that expected for a pure $d$-wave superconductor.
In the weak coupling limit the low temperature slope of the susceptibility
of a $d$-wave superconductor with $T_c=80\rm\,K$ corresponds to a
$\GdK$ shift of $10\rm\,ppm/K$\cite{gdlinprl}.

Diamagnetic corrections are not negligible for lower fields and in the
$c$ direction, and we estimate the field dependence  of $B_{\rm dia}^c$
from classical expressions and recent numerical
results\cite{Tinkham,mreversible,Ichioka_nagyPRB,SongPRB,Yaouanc_Brandt}.
Above the vortex lattice melting temperature the magnetic field
inhomogeneity due to vortices is motionally averaged and $B_{\rm dia}^c$
equals the bulk demagnetization, which is roughly proportional to
$\ln(\Hc2/B)$\cite{Tinkham,mreversible,Ichioka_nagyPRB}. Therefore one
expects a 15--30\% reduction in $B_{\rm dia}^c$ from $8$ to $15\rm\,T$.
Our bulk susceptibility measurements on the OP sample agree with such a
reduction.  A shift, $\Delta B_{\rm saddle}$, in addition to
demagnetization appears when the vortex lattice freezes, because the
magnetic field distribution is asymmetric and peaked at the saddle
point, which is different from the average field.  For zero
superconducting coherence length, $\xi_{ab}=0$, $\Delta B_{\rm
saddle}^{(0)}\approx 0.037\Phi_0/\lambda_{ab}^2\approx3\rm\,mT$ in a
triangular vortex lattice with $\lambda_{ab}=160\rm\,nm$\cite{SongPRB}.
However, numerical studies show\cite{Ichioka_nagyPRB,Yaouanc_Brandt} that
already in fields where the distance between vortices is much larger than
$\xi_{ab}$, $\Delta B_{\rm saddle}$ is reduced significantly.  Corti {\em
et al.}\cite{Corti89y} found a $3\rm\,mT$ field distribution due to
vortices in a \YBCO\ aligned powder for $B\parallel c$ at $9.4\rm\,T$
below $10\rm\,K$, and the corresponding shift is expected to be a fraction
of this value.

\begin{figure}[t]
\noindent
\includegraphics[width=\columnwidth]{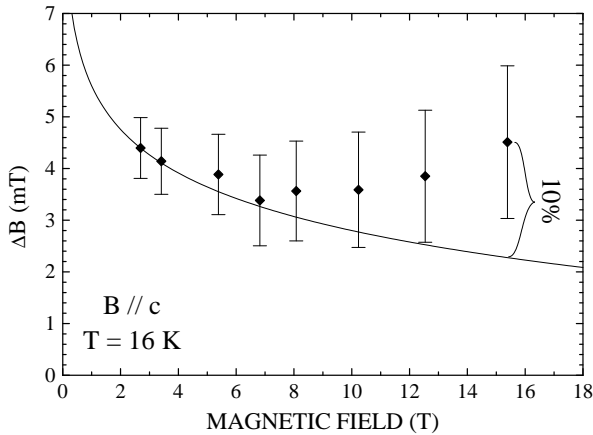}
\caption{\label{f:corr}
Field dependence of the spin susceptibility and the diamagnetic correction
at $16\rm\,K$ in absolute field scale. The solid line is a rough estimate
of $B_{\rm dia}(B,\,16{\rm\,K})$ (see text).  The error bars represent
the {\em maximum} error in the measurement. Most of the error is caused by
the uncertainty in $K_0^c-K_0^{ab}$, therefore it will be strongly
correlated between experimental points. The brace shows 10\% of the \Gd\
shift measured in $15.4\rm\,T$ at $T_c$.
}
\end{figure}

Now we estimate the increase of $\chis^c$ at $B=15.4\rm\,T$ and
$T=16\rm\,K$, the lowest temperature where experimental uncertainties are
small at all fields.
Fig.~\ref{f:corr} shows
$\Delta B=(\GdK^c-\GdK^{ab})B_0-(K_0^c-K_0^{ab})B_0$, the anisotropy of
the shifts as a function of field. In $\Delta B$ the anisotropy of
$K_0^\alpha$ is taken into account but it is uncorrected for reversible
diamagnetic effects.  The anisotropy of the ``chemical shift''
$K_0^c-K_0^{ab}=500\pm70\rm\,ppm$ was measured from the $B_0=15.4\rm\,T$
data at $0.9\,T_c$. Above this temperature a Korringa-like relaxational
broadening of the \Gd\ lines reduces precision. This value of
$K_0^c-K_0^{ab}$ is consistent with what is found in the \YBC3O{6+x}\
family\cite{gdCooper,afdomains}. Fig.~\ref{f:corr} shows that 
there is a small field dependent contribution to
$\chis^c(B_0,\,T)-\chis^{ab}(B_0,\,T) =(\Delta B-B_{\rm
dia}^c(B_0,\,T))/\GdA B_0$. $\Delta B$ varies little
with field, at higher fields it is constant or increases slightly.
In Fig.~\ref{f:corr} we illustrate the variation of the
reversible diamagnetic shift using $M_{\rm dia}(B)=M_{\rm
dia,0}\ln(\Hc2/B)$, assuming $\Hc2=100\rm\,T$ and that the field
dependence of $\chis^c$ is negligible below $3\rm\,T$, i.e., we
slightly overestimate $B_{\rm dia}$. At $15.4\rm\,T$ the shift
anisotropy is $\Delta B=4.5\pm1.5\rm\,mT$, while $M_{\rm
dia}\approx2.3\rm\,mT$.
As seen in Fig.~\ref{f:corr}, the vortex contribution to
$\chis^c(B_0=15.4{\rm\,T},\,T=16{\rm\,K})$ is of the order of 10\% of the
normal state spin susceptibility at $T_c$.

According to the semiclassical scaling relations for the spin
susceptibility derived by Kopnin and Volovik\cite{KopninVolovik}, and the
more sophisticated study of Ichioka {\em et al.}\cite{Ichioka_0.41}, in a
$d$-wave superconductor $\chis\propto(B/\Hc2)^\gamma$ with
$\gamma=0.4$ in the low temperature regime, and $\gamma\approx1$ in the
low field regime. We are close to the crossover\cite{SimonLee},
$15.4{\rm\,T}/\Hc2\approx 16{\rm\,K}/T_c$, but whatever $\gamma$ is in
our case, the 10\% enhancement implies that the spin susceptibility of the
underdoped \YBCO\ below $T_c$ is restored to $\chis(T_c)$ at maximum in a
magnetic field of $\Hc2$.
Thus the pseudogap is little affected by magnetic fields that suppress
superconductivity. Nowadays the most popular approach to explain the
microscopic origin of the pseudogap is that it is associated with
superconducting fluctuations above $T_c$\cite{preformedpair}. In a naive
picture, in which the normal state spin gap were due to incoherent Cooper
pairs, breaking of the pairs by a magnetic field would restore the spin
susceptibility to the value measured at $T\gg T_c$, which is approximately 3
times as much as $\chis(T_c)$.  Our result differs from that of Zheng {\em
et al.}\cite{ZhengClarkNMR} since they find an enhancement of the spin
susceptibility with $B$ {\em in the} $(a,b)$ plane of similar magnitude as
we do for $B\parallel c$.

In conclusion we found that at $T\ll T_c$ an applied field of
$\approx0.15\Hc2$ enhanced the spin susceptibility of the underdoped
\YBCO\ by only $\approx3$\% ($\approx10\%$) of its normal state
susceptibility measured at room temperature (at $T_c$). This suggests
that even an applied field of $\Hc2$ would not destroy the pseudogap.

We are indebted to J.~R.~Cooper for the static susceptibility measurements,
and I.~T\"utt\H o for useful discussions. Support from the US-NSF-STCS
(DMR-91-20000) and the Hungarian state grants OTKA 029150 and AKP 97-39-22
is acknowledged.


\begin{references}
\bibitem{cfparams}
The ZFS parameters of \Gd\ in Gd:\YBCO\ at $40\rm\,K$ with all the
orthorhombic terms included are
$b_2^0=-1579.2 \pm 1.5 \rm\,MHz$,
$b_2^2=-238.9  \pm 2.4 \rm\,MHz$,
$b_4^0=-188.8  \pm 0.5 \rm\,MHz$, 
$b_4^2=-11.9   \pm 5.2 \rm\,MHz$,
$b_4^4=841.9   \pm 2.9 \rm\,MHz$,
$b_6^0=0.5     \pm 0.6 \rm\,MHz$,
$b_6^2=85      \pm 45  \rm\,MHz$,
$b_6^4=-16     \pm 11  \rm\,MHz$, and
$b_6^6=-87     \pm 45  \rm\,MHz$.
(Note: the reference system of the spin Hamiltonian coincides with the
crystalline axes.)
\end{references}
\end{document}